\documentclass[twocolumn,aps]{revtex4}

\usepackage{graphicx}
\usepackage{bm}

\def\bea{\begin{eqnarray}}
\def\eea{\end{eqnarray}}

\usepackage{amssymb}
\usepackage{graphicx}

\newcommand\ovl[1]{\overline{#1}}
\newcommand {\ket}[1] {|{#1}\rangle}
\newcommand {\keta} [1] {|#1\rangle_A}
\newcommand {\ketb} [1] {|#1\rangle_B}
\newcommand {\kete} [1] {|#1\rangle_E}
\newcommand {\bra}[1] {\langle{#1}|}

\newcommand {\nil}{\emptyset}
\newcommand {\cl}{\mathcal}
\newcommand {\tsf} [1]{\textsf{#1}}
\newcommand {\norm} [1] {\parallel #1 \parallel}

\begin{document}
\title{EFFECT OF TRANSMISSION LOSS ON THE FUNDAMENTAL SECURITY OF QUANTUM KEY DISTRIBUTION}
\author{Horace P. Yuen}
\email{yuen@eecs.northwestern.edu }
\affiliation{Department of Electrical Engineering and Computer Science\\
Department of Physics and Astronomy\\
Northwestern University, Evanston, IL 60208 }
\begin{abstract}
It is shown that the effect of transmission loss has often not been properly taken into account in the security proofs on quantum key distribution. A class of general attacks to be called probabilistic re-sends attack is described that has not been accounted for, which is a generalization of the well-known unique state determination attack. In the case of the four-state single-photon BB84 protocol, it is shown in detail how such attacks are not accounted for in the known security proofs against the simplest individual attacks.
\end{abstract}
\maketitle
\section{Introduction}\label{sec:intro}
Real optical systems have significant loss from the transmitted signal to the detected signal. If the transmission loss is small one can treat the deleted bits as random errors and deal with them by an error correcting code. Security claim, however, has often been made with arbitray loss taken into account just on the throughput via post-detection selection of the detected bits. That it is clearly not a valid inference could be seen from the situation of the two-state B92, for which security is totally breached in an intercept-resend attack when the loss is above a certain threshold determined by the two signal states [1]. Such attack can be generalized to any BB84 type protocols involving any number of coherent states, which are necessarily linearly independent and hence allows such "zero error" attack in the presence of sufficient loss [2]. These general attacks have been called unique state determination (USD) attacks, and they are identical to attacks with probabilistic exact cloning [3]. For the four-state BB84, the optimal individual attack is the same as an approximate cloning attack. Probabilistic approximate cloning may, however, sometimes lead to better performance than approximate cloning itself [4]. Thus, the attacker Eve could launch attacks by probabilistic approximate cloning, and not just on single qubits (or the boson modes they embed in to be called "qumodes") but segments of qubits with entanglement or perhaps the whole key sequence. In this paper, such attack is further generalized to what will be called probabilistic re-send (PRS) attack, which has not been accounted for in the security proofs of lossy system thus far. In the case of individual attacks on the four- and six-state BB84 protocol with transmission loss, a general attack formulation is possible that include all attacks. It will be shown in the four-state case that there are indeed attacks lying outside the ones treated without loss or with loss and just throughput reduction.

\section{Incompleteness of mere post-detection selection}
The argument is often made that all the losses in the cryptosystem could just be lumped together by a loss parameter which affects the rate of final key bits that can be generated but not the key security. In [5, p.336] it is stated explicitly that "Detector inefficiencies and other types of losses can be incorporated into the Shor-Preskill security analysis easily enough. Through public discussion, Alice and Bob can eliminate from their sifted key all signals for which Bob failed to record a measurement result". In [6] only detector loss is represented with no transmission loss, presumably for the same reason since the results are applied to the NEC cryptosystem  [7,8] to operate with significant transmission loss. Although the four BB84 states are not linearly independent so that USD attack does not apply directly, a proof is needed to show why Eve cannot take advantage of the transmission loss and do better. In general, the following PRS attack needs to be fully accounted for.

Transmission loss and detector loss are very different. Assuming as we do that Bob's detector loss cannot be controlled by Eve, detector loss would delete the incoming qubits randomly independently of what Eve has done to them. On the other hand, for transmission loss Eve could intercept and then resend only those she chooses. She may or may not have the ability to replace the lossy transmission line with a lossless one. In [1-2] it was assumed she does have the capability which makes it easy to see what she may do to her advantage. Even if she does not, her attack capability is still enlarged as we will show mathematically in the case of individual attacks in Appendix A. When she has the loss replacement capability which is usually granted, we see that transmission loss becomes very different from detector loss since each of the incoming signal before detection has now been selected by Eve, which may raise her performance by an amount that has to be determined as a function of the transmission loss parameter $1-\eta$, where $\eta$ is the transmittance of the line.

\section{Probabilistic re-send (PRS) attack}
\begin{figure}
  \includegraphics[width=10cm]{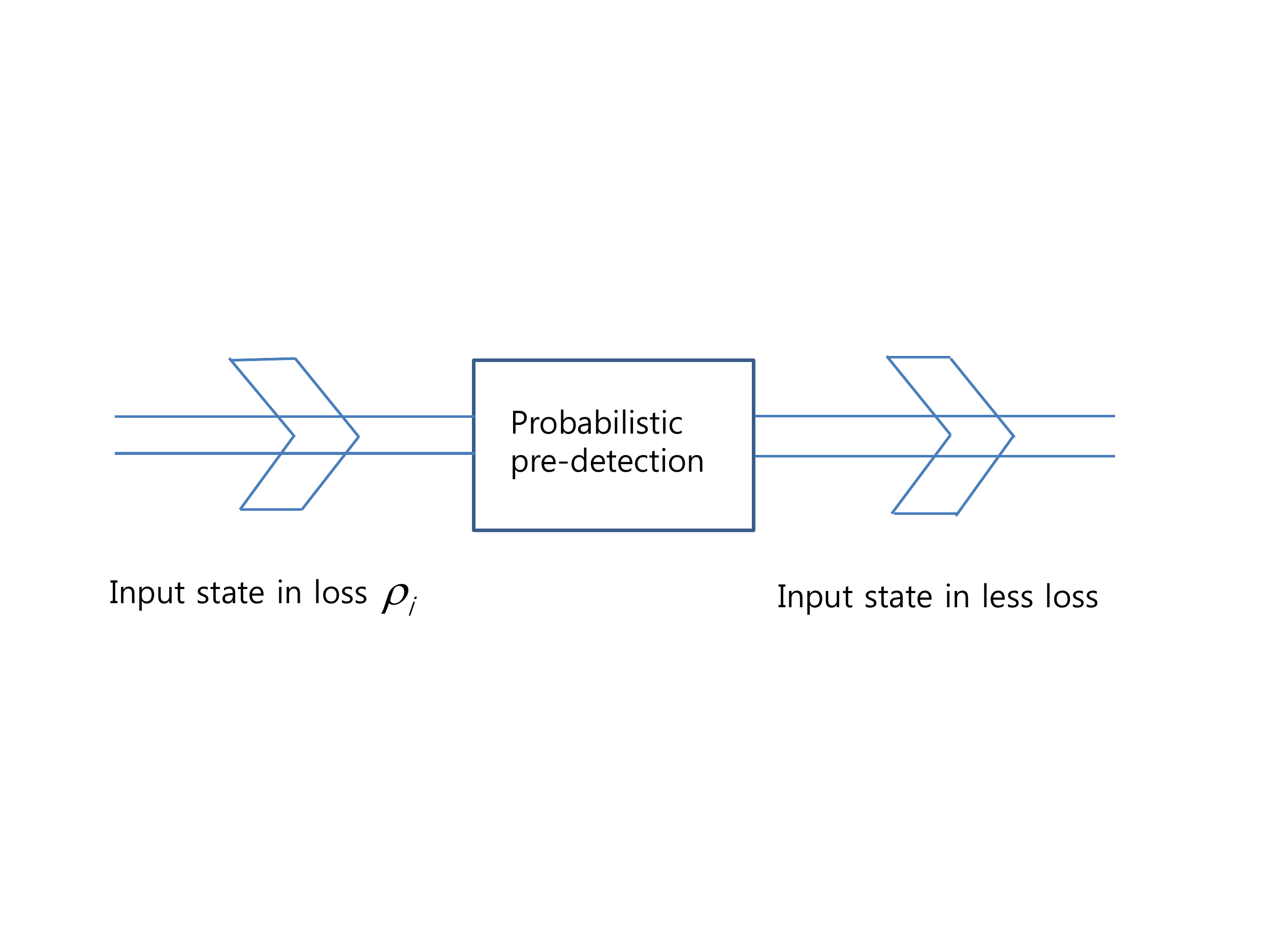}
 \caption{\label{Fig-1} Schematic way to eliminate or reduce the effect of loss by user: loss is alleviated or eliminated with favorable pre-detection outcome.}
\end{figure}

\begin{figure}
  \includegraphics[width=10cm]{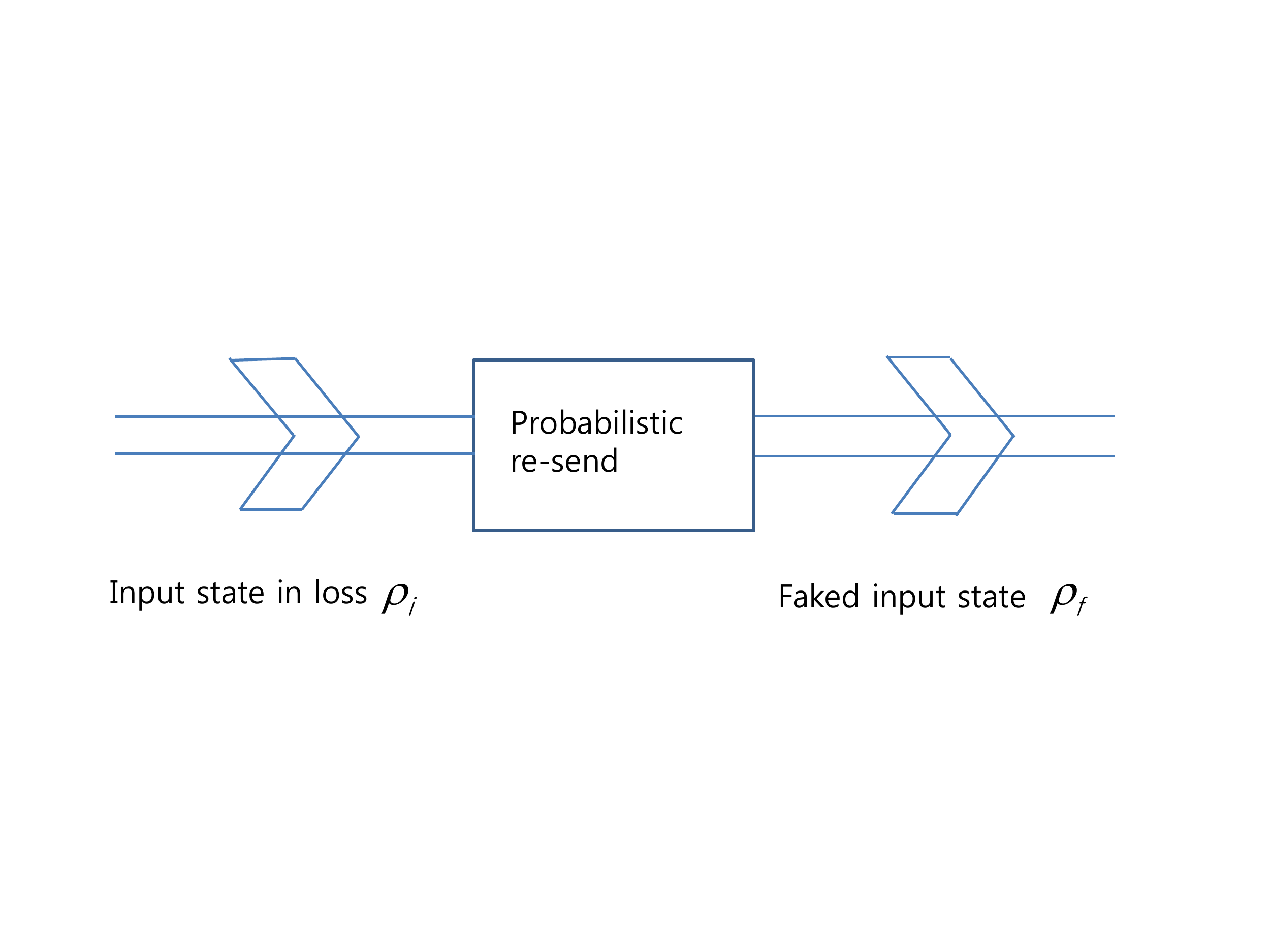}
 \caption{Schematic way to take advantage of loss by attacker: a more favorable input state $\rho_f$ from Eve's viewpoint is sought with possible quantum signal detection (PRS attack).
 \label{Fig-2}}
\end{figure}
Generally, the users may try to combat loss by pre-detection as indicated in Fig. 1, with success probability itself limited by $\eta$. Examples include quantum non-demolition   (QND) measurement (but see [9] on the inappropriate terminology) and herald qubit amplifier [10]. Eve has a similar attack approach, the probabilistic re-send attack as indicated in Fig.2. Sufficient loss would allow her to cover the deleted bits in various possible ways. The class of PRS attack evidently includes probabilistic approximate cloning. Note that this possibility of bit deletion from transmission loss {\it violates} the usual information-disturbance tradeoff that underlines QKD security of the BB84 and Ekert types which need to  employ intrusion level estimation, in that information can be gained by Eve {\it without} causing any relevant disturbance.

While PRS attacks can be covered in a sufficiently general formulation on Eve's probe formation, it is not automatically covered by mere post-detection selection as explained above, and also not by the use of squashing [11,12] or heralded qubit amplifier [10]. Specifically, squashing or QND measurement could reduce an infinite dimensional qumode to three levels, qubit plus vacuum state. Post-detection selection gets rid of the vacuum state. Heralded qubit amplifier is a pre-detection scheme of Fig. 1 that acts for entangled pairs what QND measurement does for single photons. The question remains that transmission loss may have already allowed Eve's to use PRS attack of Fig. 2 that gives her better or much better performance than a security analysis which neglects such attacks would show. Even when line replacement is not possible for Eve to make up the loss from her bit deletion, she could use PRS attacks on some fraction of the qubits allowed by the loss, which are not accounted for by mere post-detection selection. In fact, there are more attack possibilities even if line replacement is not allowed, as shown in Appendix A.

\section{Conclusion}
Significant loss cannot be avoided in optical signal transmissions. Thus, any practical application of QKD must deal with the loss induced security issues. A small amount is already known [13] to cause huge security problem when the photon detection mechanism is exploited in the detector "blinding attacks". In this paper we have shown a whole class of PRS attacks have not been accounted for in existing security analysis, not even for individual attacks when Eve cannot replace the transmission loss. Until all attacks allowed by the laws of quantum physics are taken into account, a security proof cannot be said to provide "unconditional security" even if the analysis is entirely valid.
\section*{Acknowledgement}
This work was supported by the Air Force Office of Scientific Research.
\section*{Appendix A: Identical individual attacks (IIA) on lossy single-photon BB84}
\par This appendix is derrived from a 2008 internal memo by R. Nair and H.P. Yuen.
 \par We now demonstrate that even without replacing the lossy line by a lossless one, Eve may launch attacks not covered by a lossless analysis with post-detection selection adjustment.

We make the following assumptions:
\begin{enumerate}
\item Alice's state source is perfect and prepares one of the four
BB84 states $\ket{0}$, $\ket{1}$, $\ket{+}=\frac{1}{\sqrt{2}}
(\ket{0}+\ket{1})$, or $\ket{-}=\frac{1}{\sqrt{2}}
(\ket{0}-\ket{1})$ in the two-dimensional \emph{signal space}
$\mathcal{H}_A$ in the case of the 4-state protocol. In the case
of the 6-state `BB84' protocol, the two states
$\ket{L}=\frac{1}{\sqrt{2}} (\ket{0}+i\ket{1})$ and
$\ket{R}=\frac{1}{\sqrt{2}} (\ket{0}-i\ket{1})$  are additionally
prepared. Note that, in this memo, $\mathcal{H}_A$ is simply an
abstract two-dimensional Hilbert space -- the possibility of
various implementations of this space is left open. The most
common one is that of embedding $\mathcal{H}_A$ in
$\mathcal{H}_\textsf{horiz} \otimes \mathcal{H}_\textsf{vert}$ as
the single-photon subspace, with $\ket{0}\in \cl{H}_A$
corresponding to $\ket{1}_\textsf{horiz} \ket{0}_\textsf{vert}$
and $\ket{1}$ to $\ket{0}_\textsf{horiz} \ket{1}_\textsf{vert}$.
$\mathcal{H}_\textsf{horiz}$ and $\mathcal{H}_\textsf{vert}$ are
the respectively the infinite-dimensional Fock spaces representing
the horizontal and vertical polarization modes of a field mode.

\item The received state of Bob (even in the presence of Eve) is assumed to lie in a three-dimensional Hilbert space
$\mathcal{H}_B = \mathcal{H}_A \bigoplus
\textsf{span}\{\ket{\emptyset}\}$. Here $\ket{\emptyset}$
represents the `no-count' state and is orthogonal to
$\mathcal{H}_A$ -- it is the vacuum state in the polarization
implementation described above. In the absence of Eve, the
transmission medium between Alice and Bob is represented by the
following loss map:
\begin{equation} \label{lossmap} \rho
 \mapsto \eta\rho +
(1-\eta)\ket{\emptyset}\bra{\emptyset},\end{equation} where $\rho
\in \mathcal{H}_A$ and the output state is in $\mathcal{H}_B$. In
other words, we assume (in the absence of Eve) a loss channel with
state-independent throughput (i.e., transmittance) $\eta$. It may
easily be verified that such a loss channel results in the
polarization implementation when the horizontal and vertical
polarizations are subjected to independent linear loss of
magnitude $1-\eta$.

\item Eve launches an identical individual attack (IIA). In other words, her action can
be represented by a probe system $\mathcal{H}_E$ in some initial
state $\ket{E}$ and the application of an isometry (i.e.,
inner-product preserving transformation) $T: \mathcal{H}_A \otimes
\cl{H}_E \rightarrow \cl{H}_B \otimes \cl{H}_E$ on the signal +
probe. This action is repeated with identically prepared probes on
each of the transmitted signals.

\item For each of the signal states $\ket{\psi_{in}}$, Eve's
attack results in an effective $\rho_{out} \in \cl{H}_B$. We
assume that the throughput $\eta_{\psi_{in}}:=1-
\bra{\emptyset}\rho_{out}\ket{\emptyset}= \eta$. In other words,
Eve's action results in the same throughput for each of the
four/six signal states of the 4-state/6-state protocol as would be
seen if Eve was absent. We do not make any assumption on the
throughput of the non-signal states.

\item Bob's detectors are perfect -- he is assumed to be able to
make without error and with unit success probability the ideal
measurement projecting onto $\ket{\nil}$ and any one of the
two/three orthogonal bases in the 4-state/6-state protocol.

\end{enumerate}

Our assumptions above are conservative in the sense that limiting Eve to IIA's is a restriction on her
capability. The equal signal throughput condition 4 above is also
conservative since the BB84 protocol does not explicitly include a
check for uniformity of throughput across the signal states. Even if it did, such a check would be hard to implement. In practice,
for single photon transmission through optical fibers, the throughput of the signal states may vary with time and may be polarization-dependent even in the absence of Eve. Thus, we are indeed limiting our study to a small class of attacks.
On the other hand, PRS attacks are included in the formulation when lossless line replacement is not allowed.
Unearthing new attacks in this conservative
scenario would suggest the possibility of hitherto unstudied
attacks under more general conditions.

\subsection*{A.1 Characterization of Attack Isometry $T$}

A general isometry $T: \mathcal{H}_A \otimes \cl{H}_E \rightarrow
\cl{H}_B \otimes \cl{H}_E$ is specified (at least on inputs where
$\cl{H}_E$ is in the state $\ket{E}$) by the right hand sides of
the two equations:
\begin{equation} \label{image0}
T\keta{0}\kete{E} = \sum_{i=0,1,\nil}\ketb{i}\kete{\phi_i^0},
\end{equation} and
\begin{equation} \label{image1}
T\keta{1}\kete{E} = \sum_{i=0,1,\nil}\ketb{i}\kete{\phi_i^1}.
\end{equation}
The kets of the $\cl{H}_E$ system are not normalized. The
condition that $T$ be an isometry results in the following
conditions on the states $\kete{\phi_i^b}$:
\begin{equation} \label{normcondition}
\sum_{i=0,1,\nil} \norm{\phi_i^b}^2 =1 \hspace{5mm} b=0,1,
\end{equation}
and
\begin{equation} \label{ipcondition}
\sum_{i=0,1,\nil} \bra{\phi_i^0} \phi_i^1 \rangle=0.
\end{equation}
\\
The restriction imposed by our Assumption 4 is more interesting.
By using the linearity of $T$ and the fact that the loss (loss :=
1 - throughput)  $1- \eta_\psi$ of an input state $\keta{\psi}$ is
given by the squared-norm of the state in $\cl{H}_E$ multiplying
$\ketb{\nil}$ in $T\keta{\psi}\kete{E}$, one gets the following
general expression for the loss seen by any state $\keta{\psi} =
a\keta{0}+b\keta{1}$:
\begin{equation} \label{losspsi}
1- \eta_\psi= |a|^2\norm{\phi_\nil^0}^2 +
|b|^2\norm{\phi_\nil^1}^2 + 2\tsf{Re} [a\ovl{b} \langle
\phi_\nil^0\ket{\phi_\nil^1}].
\end{equation}
\\
Let us impose Assumption 4 for the 4-state BB84 protocol. Using
(\ref{losspsi}), the condition that $\eta_0=\eta_1=\eta$ implies
that
\begin{equation}\label{direct}
\norm{\phi_\nil^0}^2 = \norm{\phi_\nil^1}^2 = 1-\eta.
\end{equation}
The restrictions that $\eta_{+} = \eta_{-} =\eta$ are both
satisfied if and only if
\begin{equation} \label{ip}
\tsf{Re}\langle \phi_\nil^0\ket{\phi_\nil^1} = 0.
\end{equation}
\\Interestingly, this condition actually implies that $\eta_\psi =
\eta$ for all $\keta{\psi} = a\keta{0}+b\keta{1}$ with $a,b \in
\mathbb{R}$, i.e., for all states on the great circle of the Bloch
sphere containing the 4 BB84 states. In the case of the 6-state
protocol, we have the additional restrictions
$\eta_L=\eta_R=\eta$. In this case, we can show from
(\ref{losspsi}) that we must have
\begin{equation} \label{ortho}
\langle \phi_\nil^0\ket{\phi_\nil^1} = 0.
\end{equation}
Adding this last condition to the rest in fact implies that
$\eta_\psi = \eta$ for all $\keta{\psi} \in \cl{H}_A$! To
summarize, the conditions (4), (5), (7), (8) must hold for both
4-state and 6-state BB84 and (9) holds for 6-state BB84.

\subsection*{A.2  Filtering of No-Count Events}

The isometry $T$ studied above contains more information than is
necessary for a security analysis. Apart from unitary freedoms in
Eve's actions and in choice of the initial probe state, we have
not yet considered the following filtering operation that Bob
performs: When Bob measures a particular signal system $\cl{H}_B$
in the state $\ket{\nil}$, that system cannot be used for
generating key and is discarded. Thus, we imagine a two-valued
projection measurement by Bob consisting of the projection onto
$\ket{\nil}$ and the projection $P_A$ onto $\cl{H}_A$. In
practice, Bob makes a single three-valued measurement, but we may
conceptually divide this into the step of making the above
two-valued measurement followed by measurement of one of the BB84
bases. Since the protocol proceeds only on the cases where Bob
obtains a result in $\cl{H}_A$, we write the post-selected output
states corresponding to inputs $\keta{0}$ and $\keta{1}$:
\begin{equation} \label{postselectedimage0}
P_AT\keta{0}\kete{E} = \sum_{i=0,1}\ketb{i}\kete{\phi_i^0},
\end{equation} and
\begin{equation} \label{image1}
P_AT\keta{1}\kete{E} = \sum_{i=0,1}\ketb{i}\kete{\phi_i^1}.
\end{equation}
We may re-normalize to norm $1$ these states by dividing by the
throughput $\sqrt{\eta}$. If we define hatted (post-selected)
states
$\kete{\hat{\phi}_i^b}=\frac{\kete{\phi_i^b}}{\sqrt{\eta}}$, we
get the following conditions on the hatted states from (4-5):
\begin{equation} \label{hatnormcondition}
\sum_{i=0,1} \norm{\hat{\phi}_i^b}^2 =1 \hspace{5mm} b=0,1,
\end{equation}
and
\begin{equation} \label{hatipcondition}
\sum_{i=0,1} \bra{\hat{\phi}_i^0} \hat{\phi}_i^1 \rangle=
-\bra{\hat{\phi}_\nil^0} \hat{\phi}_\nil^1 \rangle.
\end{equation}

For the 4-state BB84 protocol, the RHS of
Eq.~(\ref{hatipcondition}) is a pure imaginary number because of
(\ref{ip}):
\begin{equation} \label{hatip}
\langle \hat{\phi}_\nil^0\ket{\hat{\phi}_\nil^1} = iX,
\end{equation}
with $X \in \mathbb{R}$.

For the 6-state protocol, on the other hand, $\langle
\hat{\phi}_\nil^0\ket{\hat{\phi}_\nil^1}=0$. It is then readily
verified that the Eqs.~(12) and (13) are identical to the
conditions that would result from Eve applying an attack isometry
$S:\mathcal{H}_A \otimes \cl{H}_E \rightarrow \cl{H}_A \otimes
\cl{H}_E$, which corresponds to an attack on the lossless case
(this may also be seen to result from Eqs.~(2) and (3) by setting
the throughput $\eta$ identically to 1).

The equations (10) and (11) along with conditions (12) and (13)
are the starting point for the security analysis involving, for
example, the calculation of the information-disturbance tradeoff
between Eve's information and the induced error rate
[14]. From the above, we may conclude that under our
assumptions, a fresh calculation of this tradeoff is not necessary
for the 6-state protocol as it would yield exactly the same
results. For the 4-state protocol, on the other hand, the
possibility of Eve inducing a non-zero RHS in Eq.~(13) opens up a
new class of identical individual attacks that have not been
included in the security analysis to date and provide an urgent
problem for the security of the protocol.

\end{document}